# Development of EHD Ion-Drag Micropump for Microscale Electronics Cooling

C.K. Lee[1], A.J. Robinson[1,*], C.Y. Ching[2]
[1]Dept. of Mechanical and Manufacturing Engineering, Trinity College Dublin, Dublin 2, Ireland
[2]Dept. of Mechanical Engineering, McMaster University, Hamilton, Ontario, Canada, L8S 4L7IEEE Conference Publishing

*Abstract*- **In this investigation, the numerical simulation of electrohydrodynamic (EHD) ion-drag micropumps with micropillar electrode geometries have been performed. The effect of micropillar height and electrode spacing on the performance of the micropumps was investigated. The performance of the EHD micropump improved with increased applied voltage and decreased electrode spacing. The optimum micropillar height for the micropump with electrode spacing of 40μm and channel height of 100μm at 200V was 40μm, where a maximum mass flow rate of 0.18g/min was predicted. Compared to that of planar electrodes, the 3D micropillar electrode geometry enhanced the overall performance of the EHD micropumps.**

I. INTRODUCTION

With the development of advanced microfabrication techniques, Micro-Electromechanical Systems (MEMS) have shown promising potential in many real life applications. Currently there is a significant effort towards developing microfluidic systems for integrated microelectronics cooling devices. The design and functionality of micropumps play an integral role in the progress of microfluidic cooling technologies.

EHD pumping is a complex phenomenon involving the interaction between the flow field and applied electric fields [1-4]. Specifically, an ion-drag EHD micropump uses the interaction of an electric field with electric charges, dipoles or particles embedded in a dielectric fluid in order to generate a net flow [5]. In the ion-drag pumping mechanism, the Coulomb force is the main driving force, where neutral molecules get dragged along with charged ions that are moving between the electrodes (Fig. 1).

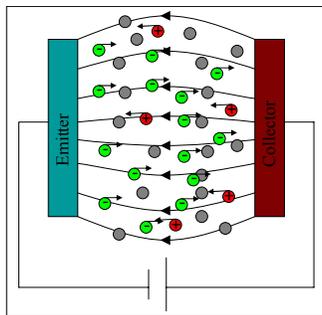

Fig.1: Ion-Drag pumping mechanism

The induced electric body force due to the interaction between the electric field and the dielectric fluid can be represented as

$$F_e = \rho_e E - \frac{1}{2}E^2 \nabla \varepsilon + \frac{1}{2}\left[\nabla E^2 \left(\frac{\partial \varepsilon}{\partial \rho}\right)_T \rho\right] \quad (1)$$

where $\rho_e$ is the charge density, $\varepsilon$ is the fluid permittivity, $\rho$ is the fluid density and T is the fluid temperature. As shown in Eq. 1 there are three components of the electric body force. The first terms is called the electrophoretic/Coulomb force and the second and third are the dielectrophoretic and electrostrictive forces.

Numerous theoretical and experimental studies have been performed on the performance and characteristics of EHD micropumps with different operating conditions, working fluids and geometrical design parameters [5-14]. The first planar EHD micropump was developed by Ahn and Kim [9]. The planar micropump was fabricated from an array of gold electrodes on a glass substrate with ethyl alcohol as working fluid. The performance of the pump was found to be highly dependent on the microchannel height.

Recent research on ion-drag micropumps has focused on optimization of micropump designs. Various ion-drag micropumps with different electrode geometries, shapes and material have been proposed [11, 12]. The potential of ion-drag micropumps in the microelectronic industries was demonstrated experimentally by Darabi and Wang [10]. Using saw tooth electrodes with an injection micro-pump, a flow rate of 3.9 g/min and a maximum pressure of 180 Pa was obtained with liquid nitrogen and HFE 7100. In a similar study, Foroughi et al. [11] reported the feasibility of ion drag micropump in cryogenic micro-cooling systems.

Darabi and Rhodes [5] simulated the 2D ion-drag pumping phenomena using a finite element method. The effects of applied voltage, electrode gap, stage gap, channel height and electrode configuration on micropump performance was modeled and validated with experimental measurements. Darabi and Rhodes [5] reported that there is an optimum spacing-to-electrode gap ratio for a given channel height, at which the flow rate is maximum. In addition, the effect of different electrode configurations was studied by modeling a microchannel with arrays of electrode on both the top and bottom walls of the channel.

## II. EHD Ion-Drag Pump with Micropillar Electrodes

To improve upon the performance of EHD micropumps with 2D planar electrodes, 3D micropillar electrodes are introduced. Micropumps with 2D planar electrode have high electric field gradient only at the sharp edges of the electrodes. With 3D micropillar electrodes, the total length of sharp edges and the total surface area of the electrode are increased significantly. As a result, the electric field gradient and the space charge density are expected to increase significantly, thereby increasing the Coulomb force and enhancing the EHD pump performance. In this study, the numerical analysis of micropumps with 3D micropillar electrodes was performed by changing (a) the applied voltage across the emitter and collector, (b) the micropillar height, $H_p$ and (c) electrode spacing between the emitter and collector, G. These are summarized in Table 1. The general design and geometry for one simulation stage of the micropump with micropillar electrodes are shown in Fig. 2. The cross section of the micropillar electrode is fixed as 40μm×40μm and the channel height, H is fixed at 100μm. The pump was first simulated under no pressure difference condition to determine its maximum mass flow rate, and subsequently set under no-flow condition to find the maximum static pressure generated.

TABLE I
TEST PARAMETERS

| Pump Design | $W_e$ (μm) | $D_p$ (μm) | $H_p$ (μm) | G (μm) | S (μm) | $V_{exp}$ (V) |
|---|---|---|---|---|---|---|
| G40S80 | 40×40 | 60 | 60 | 40 | 80 | 100-300 |
| G40S80 | 40×40 | 60 | 20-100 | 40 | 80 | 200 |
| GxxSxx | 40×40 | 60 | 60 | 30-50 | 60-100 | 200 |

## II. Computational Approach

A 3D numerical model was performed using commercial Finite Element package Cosmol Multiphysics to model the EHD pumping between one emitter and collector stage. Fig. 2 depicts a schematic diagram of the physical model used in the simulations. The governing equations are the mass, momentum and charge conservation equations;

$$\frac{\partial \rho_m}{\partial t} + \frac{\partial (\rho u_j)}{\partial x_j} = 0 \quad (2)$$

$$\frac{\partial (\rho u_i)}{\partial t} + \frac{\partial (\rho u_i u_j)}{\partial x_j} = -\frac{\partial P}{\partial x_i} + \frac{\partial \tau_{ij}}{\partial x_j} + \rho f_i + \rho_e E \quad (3)$$

$$\frac{\partial \rho_e}{\partial t} + \nabla \cdot J = 0 \quad (4)$$

where $\rho_e E$ is the Coulomb force and J is the current flux. The electric field can be related to the potential field, Φ, by the expression,

$$E = -\nabla \phi \quad (5)$$

The current flux is composed of three components, (i) flux due to convection of charge, (ii) flux due to diffusion and (ii) flux due to ionic mobility, and may be written as:

$$J = \rho_e u + \sigma E + \mu_e \rho_e E \quad (6)$$

where $\rho_e$, $\sigma$ and $\mu_e$, are the unipolar charge density, conductivity of the bulk liquid and mobility coefficient respectively. Combining Equations (4) and (6) for steady state gives:

$$\nabla \cdot (\rho_e u) + \nabla \cdot (\sigma E) + \nabla \cdot (\mu_e \rho_e E) = 0 \quad (7)$$

The three components are the convective charge current, electrical conduction current and migration charge current respectively. Finally, Gauss's Law provides the relationship between charge density and electric field, and is given by,

$$\nabla \cdot (\varepsilon E) = \rho_e \quad (8)$$

Several assumptions and approximations are made during the modeling and solution of the governing equations for the ion-drag pump. The assumptions and simplifications are:

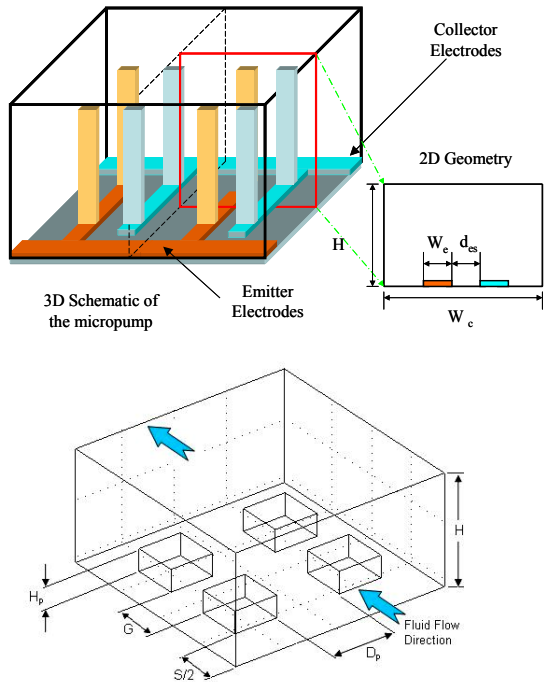

Fig. 2: Schematic of micropillar pump

laminar, steady and incompressible flow of a Newtonian fluid, only unipolar charges, the Coulomb force is the only current flux which contributes to the flow field; ion migration effect is dominant over the diffusion of charges, the side wall and entrance effects were ignored in the flow field modeling, only a one-way coupling between the Navier-Stokes equation and conservation of charge equation, and the conductivity component of conservation of charge equation is assumed negligible

The boundary condition settings for electrostatics (Gauss's Law) are straight forward. The surface of the emitter is set according to the applied voltage while the collector is grounded. Other surfaces were specified as electrically insulated. To characterize the periodicity of the micropump, the inlet and outlet boundaries are imposed with periodical boundary condition. The setting of boundary conditions for solving the conservation of charge equation is a major challenge for numerical modeling of EHD pumping as the mechanism of ion injection from emitter electrode to working fluid is still unresolved. To date, there is no complete understanding on charge generation on the emitter. The solution employed here was to impose a Dirichlet boundary condition and set the charge density as the value obtained from the experimental electric current [16]

$$\rho_e = \rho_{e,emitter} \text{ on the emitter so that } I_{exp} = \int_A J \cdot dA \quad (10)$$

For the flow field equations all surfaces except the inlet and outlet are set to a no-slip condition. The inlet and outlet are specified accordingly to the desired flow conditions of the pump. Two configurations of micropump were studied, (i) closed loop to obtain the maximum pressure generation and, (ii) opened loop to determine the steady flow rate achieved with the micropump. In the first instance, the pressure at the inlet was set to zero (atmospheric pressure) while the outlet was specified as a no-slip wall. From this setting, the maximum pressure head generated can be obtained. In the second case, both inlet and outlet of the pump were set to be atmospheric pressure, hence no pressure difference between these two points. A steady state maximum flow velocity can be determined from this configuration.

### III. RESULTS AND DISCUSSION

Figure 3 shows an example of the simulation results for pump G40S80_Hp60 with an applied voltage of 200V for the open-loop flow boundary conditions. The working fluid is HFE-7100 and the channel height is 100 μm. The figure shows the Coulomb force, cross-sections of the pressure field and a vector plot of the velocity field. Fig. 3a indicates that the strongest driving force occurs between the electrodes, as is expected and consistent with simulations for 2D planer electrodes [5]. Relative pressures (Fig. 3b) are thus highest between the micropillar pairs which affect the local velocity field as depicted in Fig. 3c. As discussed by Darabi and Rhodes [5], the electric body force, acting on the ions, drags the liquid molecules between the electrodes. The regions of high velocity thus concentrated between the electrodes where the electric field influence on the liquid is strongest

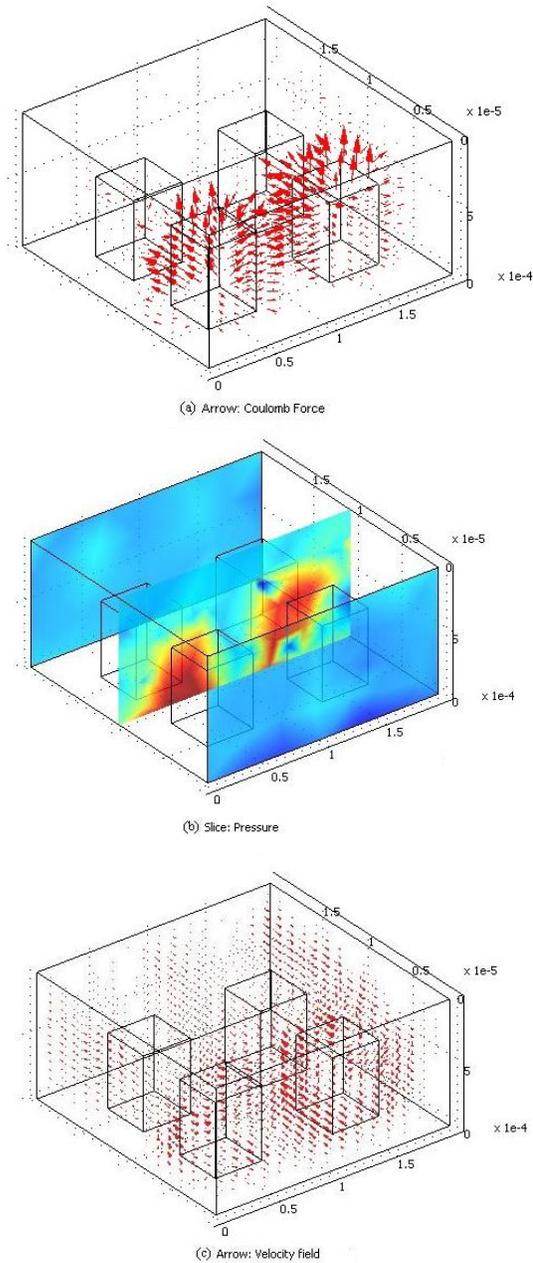

Fig. 3: Pump G40S80_Hp60 with an applied voltage of 200V: (a) Coulomb force, (b) Pressure and (c) Velocity field

*A. Effect of Applied Voltage*

The effect of applied voltage on the EHD micropump performance was studied for one stage of the micropump with two pairs of 60μm high 3D micropillar electrodes. The distance between the emitter and collector (G) is 40μm while the distance between two electrodes ($D_p$) with same polarity is fixed as 60μm. Simulations were performed for applied voltage from 100V-300V.

The effect of applied voltage on the static pressure under no-flow condition and the mass flow rate under no pressure difference are summarized in Fig. 4. Both the static pressure and mass flow rate increase as the applied voltage is increased. In addition, the rate of increase of both parameters increases with applied voltage.

The numerical simulations results suggest that the performance of the EHD micropump in terms of static pressure and mass flow rate for a certain pump design can be improved or manipulated accordingly to meet various requirements of different applications. Adjustment of applied voltage is a relatively simple and effective method to control the characteristic of EHD micropump.

the pump domain.

In this analysis, pump G40S80 is analysed with an applied voltage of 200V. The height of the micropillar was varied from 20μm to 100μm, and the static pressure generated under no-flow condition and the mass flow rate of working fluid with no pressure differential are shown in Fig. 5. Under no-flow condition, the pump G40S80 with applied voltage 200V recorded a large increase in static pressure from 70 Pa to 250 Pa, as the micropillar height increased from 20μm to 60μm. As the micropillar height is further increased from 60μm, the increase rate of static pressure drops significantly and eventually reaches a maximum pressure value of 275 Pa at micropillar height of 80μm. The static pressure decreases slightly when the micropillar height increases beyond 80μm. This result is expected as the rise in micropillar height would greatly increase Coulomb force within the flow domain when the micropillar initially increased from 20μm to 60μm. The increase in Coulomb force is significant compared to increase in flow resistance at this early stage of analysis. Further increase in micropillar height would be followed by a higher increase in the rate of flow resistance and eventually cancel out the effect of increases in Coulomb force

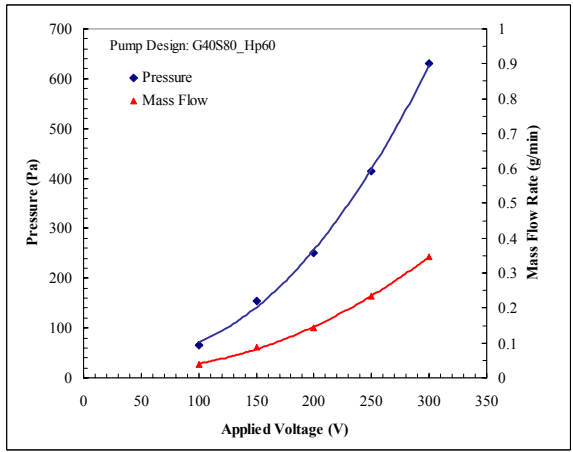

Fig. 4: Effect of applied voltage on static pressure and mass flow rate for pump G40S80_Hp60.

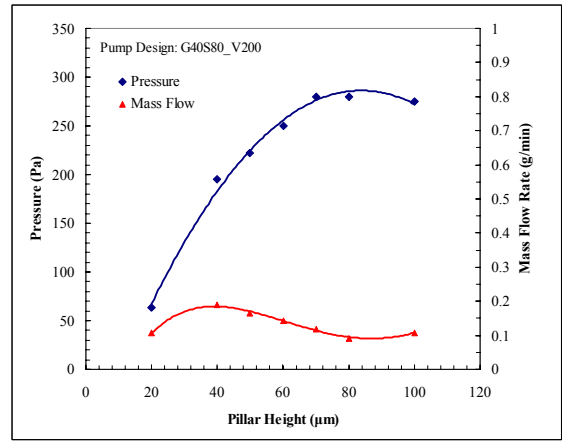

Fig.5: Effect of micropillar height on static pressure and mass flow rate for pump G40S80 with applied voltage of 200V under no-flow condition.

*B. Effect of Micropillar Height*

An increase in the micropillar electrode height results in a significant increase of the overall surface area and the length of sharp edges along the emitter. The Coulomb force within the pump domain is expected to increase due to the higher electric field gradient and space charge density. However, an increase in micropillar height at the same time increases the resistance that impedes the fluid flow. Therefore, the effects of micropillar electrode height on the micropump performance are governed by the mutual interaction between the increased Coulomb force and the flow resistance within

The mass flow rate of the pump is highly affected by the flow resistance induced by micropillar electrodes. As illustrated in Fig.5, the mass flow rate increased to the maximum value of 0.18 g/min as the micropillar height is increased from 20μm to 40μm. As the micropillar height further increased, the mass flow rate decreases steeply until the minimum value of 0.09 g/min at a micropillar height 85μm before rising again slightly as the micropillar height reaches the channel ceiling of 100μm. This result can be explained with the relationship and interaction between the Coulomb force and flow resistance. When the micropillar height is increased from 20μm to 40μm, the raise in Coulomb force is more significant relative to the increase in flow resistance until the maximum point of mass flow rate where the flow resistance begins to partially offset the effect of increasing the Coulomb force. The flow resistance has a more significant effect on the fluid flow field for micropillar height of 40μm to 60μm. The effect of Coulomb force once again overcomes the effect of flow resistance and the mass flow rate increased slightly when the micropillar height reaches the ceiling level of the channel.

C. *Effect of Electrode Spacing*

In the study of the effect of the electrode spacing on micropump performance, a micropillar electrode height of 60μm is used and a potential voltage of 200V is applied across the emitter and collector. Three pump designs with electrode spacing of 30μm, 40μm and 50μm, namely G30S60_Hp60, G40S80_Hp60 and G50S100_Hp60 are modeled and analyzed.

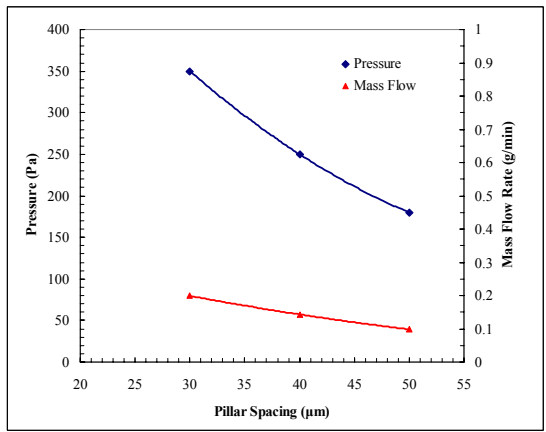

Fig.6: Effect of electrode spacing on static pressure and mass flow rate for pump with micropillar height of 60μm and an applied voltage of 200V under no-flow condition

The effect of electrode spacing on the static pressure under no-flow condition and mass flow rate under no pressure difference are shown in Fig. 6. Both static pressure and mass flow rate increased as the electrode spacing between emitter and collector is decreased. Under no-flow condition, the static pressure recorded an improvement from 180 Pa to 350 Pa as the electrode spacing decreased from 50μm to 30μm. Under no pressure difference condition, the mass flow rate increased from 0.10 g/min to 0.20 g/min for the same alteration in electrode spacing. This observation can be explained by the fact that the electric field gradient is inversely proportional to the spacing between two electrodes with opposite polarity. In other words, the electric field gradient is higher for the pump with smaller electrode spacing compared to pump with higher electrode spacing

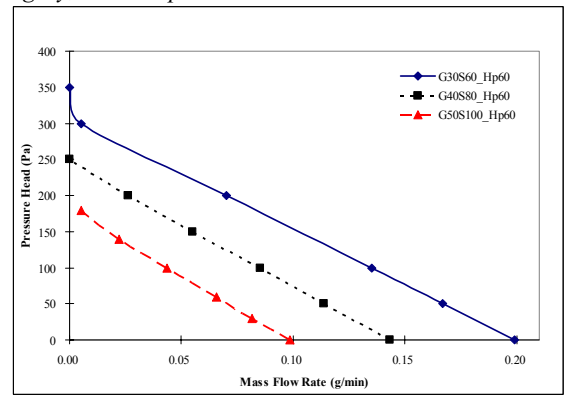

Figure 7: Characteristic curves for micropumps with different electrode spacing, 60μm micropillar height and at 200V applied voltage.

The pump characteristic curves for micropumps with three different electrode spacing range from 30μm to 50μm with micropillar height at 60μm and at an applied voltage of 200V is shown in Fig. 7. The pump with the lowest electrode spacing (G30S60_Hp60) achieved significantly higher static pressure and mass flow rate. Hence, the operating range for this micropump is noticeably wider compared to the other two pump designs. The large operating range for micropumps with small electrode spacing implies higher feasibility of the pump in meeting the requirements in various applications.

D. *Planer vs. Micropillar Electrodes*

A parallel numerical investigation of 2D planer electrodes was carried out. A comparison of pressure head and mass flow rate achieved for a micropump with 2D planar electrodes and 3D micropillar electrodes, with same electrode length (40μm×40μm) and electrode spacing of 40μm are shown in Fig.8. In the comparison of mass flow rate under no pressure difference, the micropumps with 2D planar electrodes were assumed to have a channel width (in direction normal to simulation domain) of 4mm, which were the same as width of micropumps with 3D micropillar electrodes.

The result show that the micropump with 3D micropillar electrodes required a considerably lower applied voltage to achieve a specified pressure head or mass flow rate, compared to micropump with 2D planar electrodes. To generate a pressure head of 500 Pa, the micropump with 2D planar electrodes (G40S80) required an applied voltage of 800V while micropumps with 3D micropillar electrodes (G40S80_Hp60) only required an applied voltage of 275V. Using the same pump design, a mass flow rate of 0.35 g/min is achieved with micropump with 3D micropillar electrodes at 300V, while the micropump with 2D planar electrodes required a voltage of 650V to achieve the same mass flow rate.

These numerical results seem to confirm that the proposed 3D micropillar electrodes managed to improve the micropump performance significantly. The substantial improvements in pump performance are due to the significant increased in total electrode surfaces and length of sharp edges, and hence enhanced the charge injection, and increased the electric field gradient.

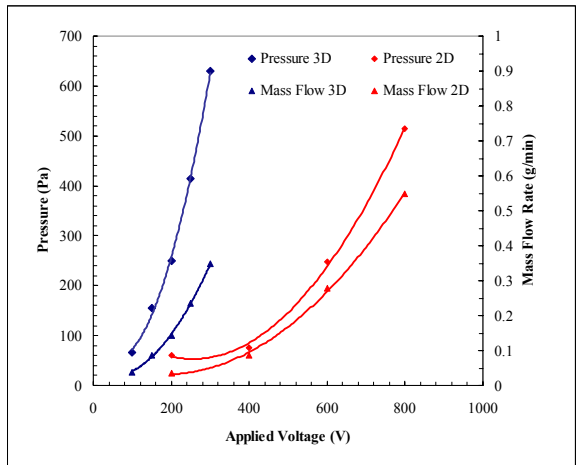

Fig.8: Comparison of pressure head generated and mass flow rate obtained at different applied voltage for micropump with 2D planar electrodes and 3D micropillar electrodes.

## IV. CONCLUSIONS

Parametric numerical studies on the 3D micropillar electrodes were performed and the results have been very promising. The performance of the EHD micropump improved with an increase in the applied voltage and a decrease in the electrode spacing. The optimum micropillar height for the micropump with electrode spacing of 40μm and channel height of 100μm at 200V, under no pressure difference across the pump inlet and outlet was 40μm, where a maximum mass flow rate of 0.18g/min was obtained. Compared to that of planar electrodes, the 3D micropillar electrode geometry enhanced the overall performance of the EHD micropumps.